\documentclass[final,3p,times,twocolumn]{elsarticle}

\usepackage{amssymb}
\usepackage{amsmath}

\usepackage{upgreek}
\usepackage{xcolor}
\usepackage{ulem}

\usepackage{array}
\newcolumntype{C}[1]{>{\centering\arraybackslash}p{#1}}

\journal{OFT}

\begin{document}

\begin{frontmatter}

\newcommand\blfootnote[1]{%
  \begingroup
  \renewcommand\thefootnote{}\footnote{#1}%
  \addtocounter{footnote}{-1}%
  \endgroup
}



\title{Polarization Maintaining Large Mode Area Yb Fibers for All Fiber Nanosecond Pulse Amplification} 


\author{Erin S. Lamb\corref{cor1}\fnref{label1}}

\ead{erin.lamb@lightera.com}
\cortext[cor1]{}

\author{Yaakov Glick\fnref{label2}}
\author{Jose Pincha\fnref{label1}} 
\author{Ishu Goel\fnref{label1}}
\author{Robert S. Windeler\fnref{label1}}
\author{Simona Ovtar\fnref{label3}}
\author{Vasiliy Lukonin\fnref{label1}}
\author{Ian Sun\fnref{label1}}
\author{Shantanu Pandit\fnref{label1}}
\author{and Jeffrey W. Nicholson\fnref{label1}} 

\affiliation[label1]{organization={Lightera},
            addressline={19 Schoolhouse Road}, 
            city={Somerset},
            postcode={08873}, 
            state={NJ},
            country={USA}}

\affiliation[label2]{organization={Soreq NRC},
            addressline={Applied Physics Division}, 
            city={Yavne},
            postcode={8180000}, 
            country={Israel}}

\affiliation[label3]{organization={Lightera},
            addressline={Priorparken 680}, 
            city={Brøndby},
            postcode={2605}, 
            country={Denmark}}

\begin{abstract}

Polarization maintaining (PM), all-fiber amplifiers offer the benefits of alignment free and environmentally stable operation. To achieve high output powers, particularly in pulsed operation, it is necessary to balance the need to reduce deleterious nonlinear effects, often through the use of large mode area (LMA) fibers, with the onset of transverse mode instability whereby higher order modes (HOMs) mix with the desired fundamental mode output. Over the last few years, advances in high HOM loss, ytterbium-doped LMA fibers have enabled continuous wave (CW) output powers up to 5~kW and pulse energies up to 2~mJ in non-PM fibers. In CW operation, LMA PM fibers have shown up to 2~kW of average power. In this contribution, we present all-fiber nanosecond pulsed amplification in a high HOM loss, Yb-doped LMA fiber with a 26~\,$\upmu$m mode field diameter and 2.4~dB/m of pump absorption at 976~nm, achieving 1~mJ of pulse energy at 1~kW of average power, and 1.5~mJ of pulse energy at 750~W of average power. The polarization extinction ratios were 20~dB or higher and the M$^{2}$ was near the diffraction limit. We measured the in-pulse to out-of-pulse energy and found 99.9\% or more of the measured power remained in-pulse.
\end{abstract}




\end{frontmatter}



\section{Introduction}
\label{sec1}

High power fiber lasers and amplifiers have numerous applications, including micromachining, defense, scientific research, advanced light detection and ranging (LIDAR), gravity wave detection, and quantum computing. Fully fiber-integrated systems are desirable for their alignment stability over their solid-state counterparts, and, further, polarization maintaining (PM) systems provide enhanced environmental stability that benefits most applications. For example, the stability advantages of all-fiber, PM sources make them advantageous for machining in settings with large vibrations or temperature fluctuations that can disrupt non-PM systems \cite{Liu2025} and for vehicle-deployed frequency combs, such as for oil and gas well monitoring \cite{Sinclair2014} or space-based operation \cite{Lezius2016,Probster2021}. Coherent beam combination \cite{Fathi2021,Klenke2018,Boland2025}, coherent LIDAR \cite{Cariou2006}, and nonlinear frequency conversion systems \cite{Andrekson2020,Hansryd2002} also rely on polarization stability and thus require PM or polarization controlled sources.

\begin{figure}
\centerline{\includegraphics[width=3.4in]{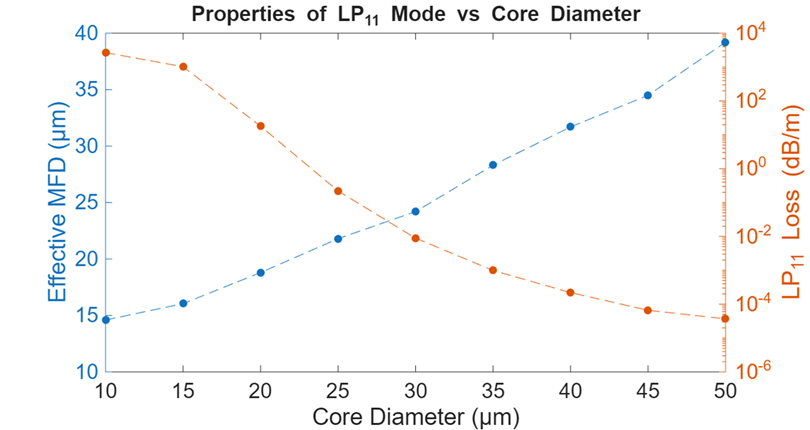}}
\caption{Effective mode-field diameter and LP$_{11}$ loss as a function of core diameter for a step-index fiber operating at 1070~nm wavelength and coiled to 15~cm bend diameter.}
\label{fig:MFDvsCoreDiameter}
\end{figure}

Nonlinear effects, such as stimulated Brillouin scattering (SBS), stimulated Raman scattering (SRS), and self-phase modulation (SPM), limit the output power and energy from fiber laser systems. The nonlinear thresholds of the fiber can be increased by increasing the mode-field diameter (MFD) of the fiber. However, the output beam quality from larger-core fibers quickly degrades due to the onset of transverse mode instability (TMI), whereby the desired fundamental mode output interferes with the higher order modes (HOMs) supported by the fiber, leading to a temporally unstable output beam \cite{Jauregui2020}. The magnitude of this issue is demonstrated in Fig. \ref{fig:MFDvsCoreDiameter}, which shows the effective MFD and loss of the LP$_{11}$ mode with a 15~cm coil as a function of core diameter as calculated with a mode solver \cite{Fini2006}. The loss of this HOM rapidly decreases with increasing fiber core diameter. For example, the HOM loss of a 30~\,$\upmu$m core is less than 0.01~dB/m and such a fiber is effectively few-moded over lengths relevant to a fiber laser. Thus, to be successful in reducing detrimental nonlinear effects with LMA fibers, it is important to simultaneously increase HOM losses over conventional fiber designs or otherwise manage HOM content to increase the TMI threshold. 

In addition to HOM loss enhancement, other strategies are available for increasing the TMI and nonlinear thresholds of LMA fibers. These include adjusting the material composition of the fiber to tailor the thermo-optic and Brillouin coefficients to achieve over 1~kW of continuous wave (CW) power \cite{Hawkins2021}. The effect of material composition on these thresholds and projected laser performance has also been simulated \cite{Mermelstein2025}. In another approach, photonic bandgap fibers use engineering of the glass structure to support single-moded operation and have achieved 1.37~kW of output CW power \cite{Pulford2021}.

\begin{figure}
\centerline{\includegraphics[width=3.5in]{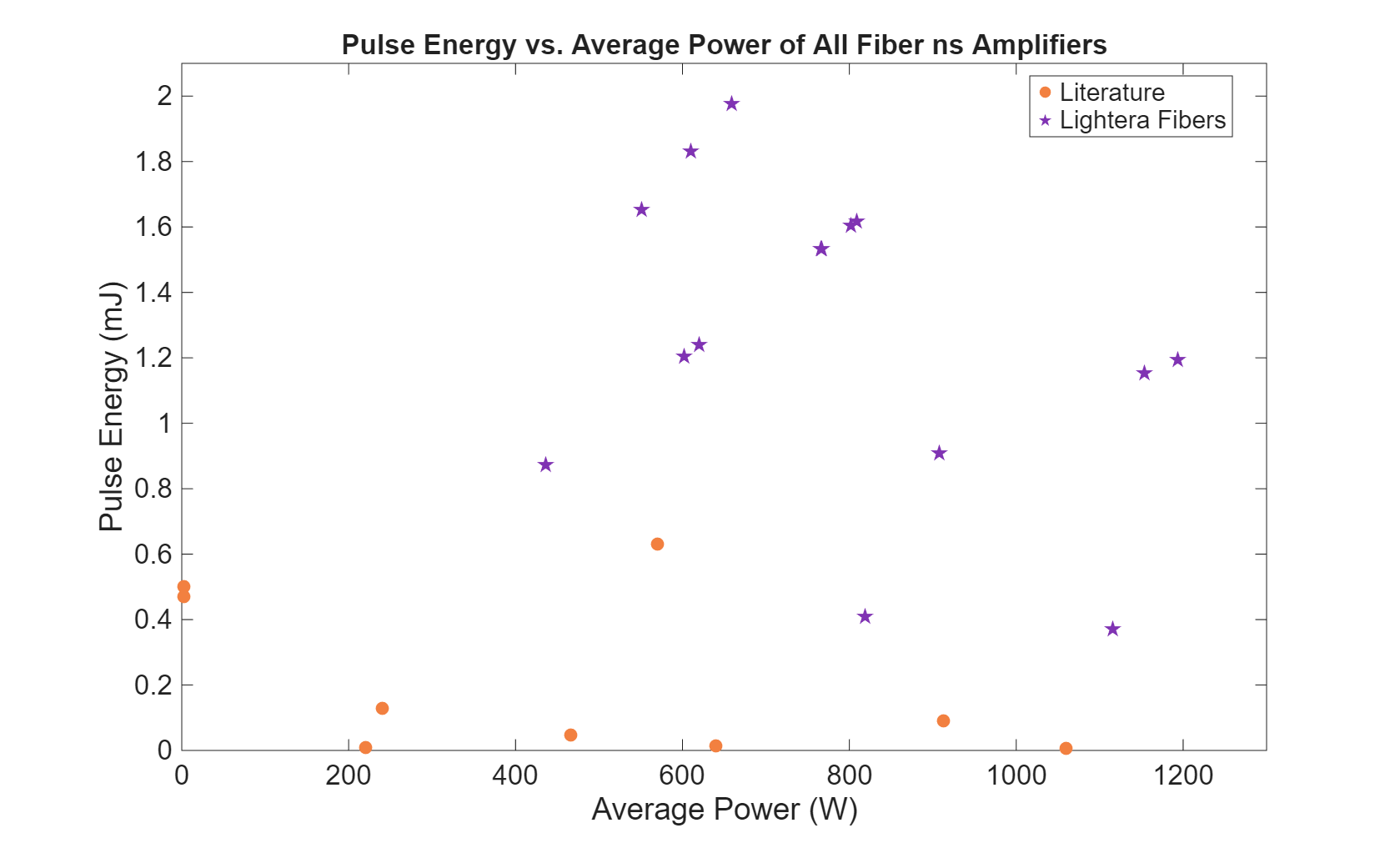}}
\caption{All-fiber, diffraction limited, nanosecond reports of pulse energy vs. average power of non-PM fibers. Literature data points are from Refs. \cite{Zhang2017,Su2014,Huang2018,Fang2011,Yu2014,Fu2018,Avdokhin2015}. Figure adapted from Ref. \cite{Glick2024}.}
\label{fig:NonPMLitSummary}
\end{figure}

A new generation of high HOM loss LMA Yb fibers are enabling kilowatts of output power, and, in nanosecond pulsed operation, simultaneous pulse energies over a millijoule. In CW operation, Yb 22/400 fibers with an HOM loss 3x that of conventional fiber designs have been found to support 5~kW of TMI-free output power \cite{Nicholson2023}. Extending these results to larger 25 to 26~\,$\upmu$m MFDs and higher absorption fibers to support pulsed operation with lower nonlinearities, over 2~mJ of pulse energy with 660~W of average power and and over 420~kW of peak power and 1.6~mJ pulses with 370~W of average power and over 1.2~kW of peak power have been demonstrated \cite{Glick2024}. The results of these fibers are shown relative to prior published results on all-fiber nanosecond amplifiers in Fig. \ref{fig:NonPMLitSummary}, which shows the advantage these new fiber designs have in achieving both high pulse energy and high average power in this regime. 

\begin{table*}[ht]
            \caption{Lightera Large Mode Area Fiber Properties. Line 7 represents the PM 30/400 fiber described in this manuscript. Further details available in Ref. \cite{Glick2024}. B: birefringence.}
            \centering
            \begin{tabular}{|C{0.6cm}|C{1cm}|C{1.1cm}|C{1.1cm}|C{1.1cm}|C{1.1cm}|C{1cm}|C{1.1cm}|C{1.4cm}|c|}
                \hline
                \ PM & Core Dia. ($\upmu$m) & MFD @ 1064 nm ($\upmu$m) & Abs. @ 976 nm (dB/m) & HOM Loss Rel. to Fiber in Ref. \cite{Nicholson2023} & B & TMI Limit (kW) & Fiber Length (m) & Result & Ref\\
                \hline
                N & 20 & 19.4 & 1.4 & 0.3 & - & 3.2 & 9 & 3.2 kW  & \cite{Nicholson2022,Nicholson2023}\\
                \hline
                N & 19.1 & 19.8 & 1.8 & 1 & - & 5 & 7 & 5 kW & \cite{Nicholson2023}\\
                \hline
                N & 21.4 & 20.1 & 2.3 & 0.37 & - & 3.3 & 3.6 & 0.87 mJ & \cite{Glick2023}\\
                \hline
                N & 25 & 24.7 & 2.4 & 0.75 & - & 2.2 & 5 & 2.2 kW & \cite{Glick2023}\\
                \hline
                N & 31.2 & 24.8 & 1.9 & 0.88 & - & 1.8 & 3.2 & 1.6 mJ & \cite{Glick2023ASSL}\\
                \hline
                N & 41.6 & 26.3 & 4.5 & 0.5 & - & 1 & 2.5 & 2 mJ & \cite{Glick2024SPIE}\\
                \hline
                Y & 30 & 26 & 2.4 & 0.9 & 1e-4 & 2 & 4.6 & 1.5 mJ & New \\
                \hline
            \end{tabular}
            \label{Fibers}
        \end{table*}

Enhanced HOM loss PM fibers allow for improved environmental stability at average powers of over 1~kW. Using a 30/400 ytterbium-doped fiber with octangonal shaped cladding and a 26~\,$\upmu$m MFD and 2.4~dB/m of pump absorption at 976~nm, CW operation with up to a TMI limit of 2~kW along with a polarization extinction ratio (PER) over 20~dB and an M$^{2}$ of 1.1 was demonstrated \cite{Nicholson2025}. Preliminary few-nanosecond pulsed amplification experiments from this fiber have achieved 0.86~mJ of energy at 430~W of average power \cite{Lamb2025} and 0.5~mJ at 1~kW of average power \cite{Lamb2025CLEOEurope}. Table \ref{Fibers} compares this fiber to previous LMA, high HOM loss non-PM fibers. Other techniques for power and energy scaling in PM fibers involve using tapered fiber structures to support hundreds of Watts of output power from PM amplifiers in CW operation \cite{Roy2024} and in picosecond pulse operation \cite{Fathi2024}. Depressed cladding fiber designs have also been shown to support up to 400~W of CW output power \cite{Kruska2024}. 

\begin{figure}
\centerline{\includegraphics[width=3.5in]{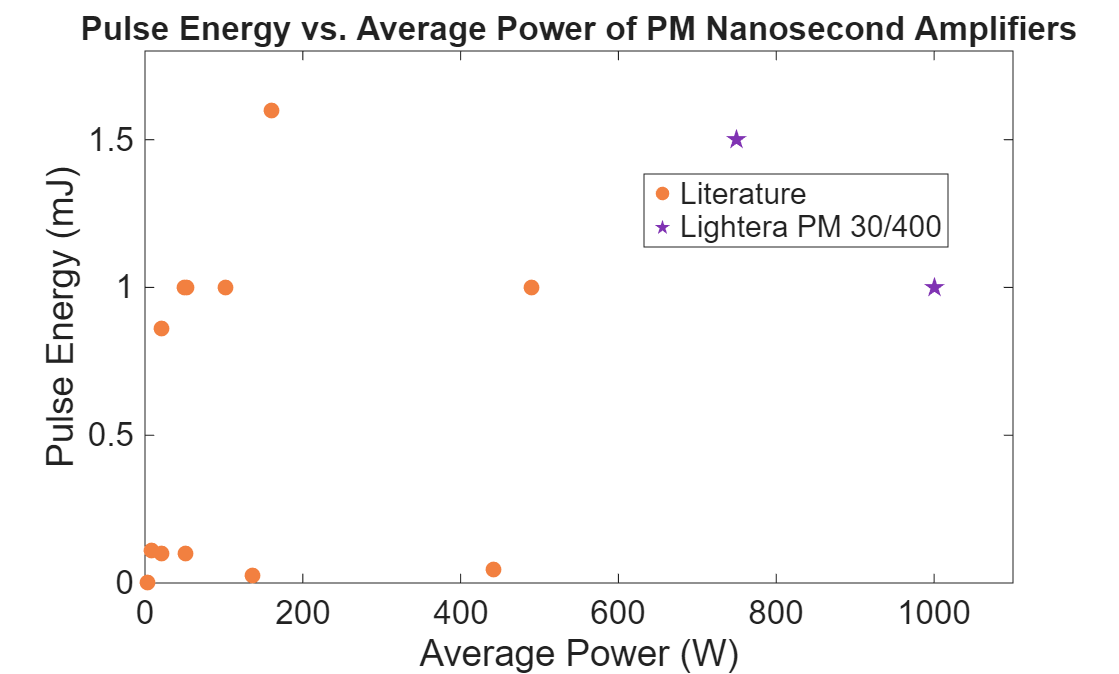}}
\caption{Pulse energy vs. average power for published nanosecond PM Yb fiber amplifiers in the literature compared with the Lightera PM 30/400 fiber featured in this work. Literature results are taken from Refs. \cite{Khitrov2008,Ye2006,Lin2017,huang2021,Su2013,Huang2018,Ran2015,Su2019,zhang2013,Roy2021,Fathi2025}.}
\label{fig:PMLitGraph}
\end{figure}

In this current work, we show 1~mJ of pulse energy from a high HOM loss PM 30/400 fiber, which yielded 1~kW of average power at a 1~MHz repetition rate. The system is all-fiber. The PER averaged 23.5~dB, the M$^{2}$ was around 1.11, and we measured that over 99.94\% of the power resides in the main pulse. We also achieved 1.5~mJ of pulse energy at 500~kHz, which gave 750~W of average power, along with a 20~dB PER and an M$^{2}$ of 1.13. 99.9\% of the power remained in-pulse. In Fig. \ref{fig:PMLitGraph}, these results are shown relative to nanosecond Yb-doped PM fiber amplifiers published in the literature, again indicating the combination of high average power and high pulse energy achievable with these designs.  

\section{Experimental Setup}

\begin{figure*}
\centerline{\includegraphics[width=6.5in]{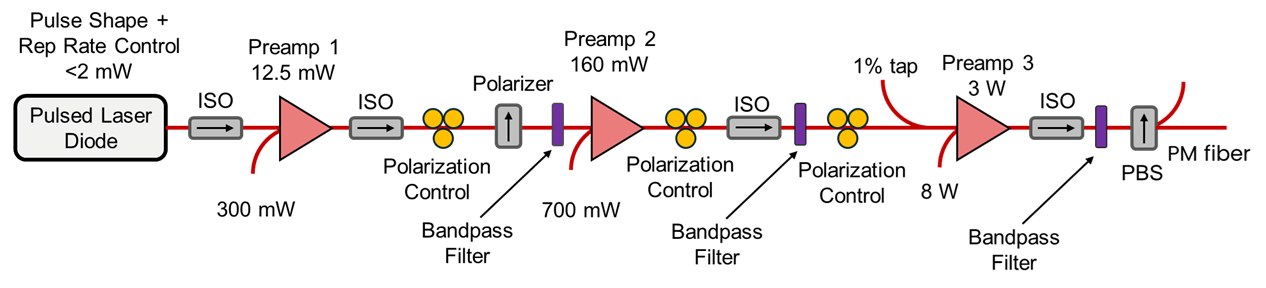}}
\caption{Schematic of seed source for power amplifier, consisting of a pulsed laser diode and three stages of pre-amplification. ISO: isolator; PBS: polarizing beam splitter; PM: polarization maintaining.}
\label{fig:SourceSetup}
\end{figure*}

The flexible seed source used in the amplification experiments is depicted in Fig. \ref{fig:SourceSetup}. It started with a gain switched diode seed laser with tunable repetition rate and pulse duration (Aerodiode 1064LD-1-5-1) that emitted less than 2~mW of power with the parameters used in these experiments. After an isolator, the pulse was next amplified to 12.5~mW in 120~cm of core-pumped, single mode fiber (ThorLabs Yb1200-4/125). After an additional isolator, a polarization controller was used to maximize transmission through a polarizer, and a bandpass filter was used to mitigate amplified spontaneous emission (ASE) and control spectral bandwidth. A second 2~m core-pumped, single mode preamplifier (ThorLabs Yb1200-4/125) then boosted the pulse to around 160~mW of average power. An isolator, two polarization controllers, and an additional bandpass filter were used prior to a third preamplifier constructed with 2.1~m of double clad fiber (Lightera Yb10-130). The output power after this preamplifier was 3~W, and the preceding polarization controllers were used to maximize the output power through a polarizing beam splitter prior to the PM power amplifier. An isolator and bandpass filter were also used between this preamplifier and the power amplifier. 

\begin{figure*}
\centerline{\includegraphics[width=5.5in]{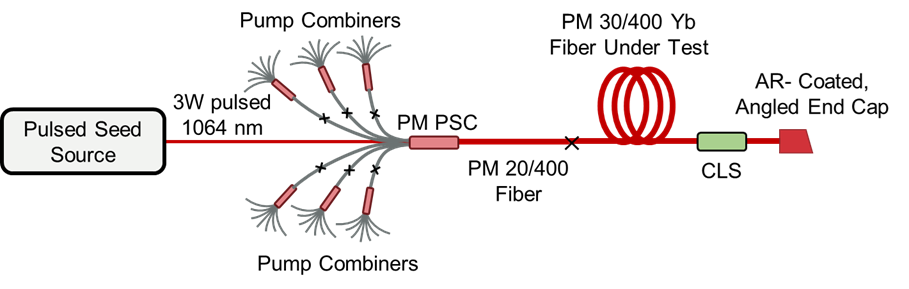}}
\caption{Schematic of power amplifier, consisting of the seed source shown in Fig. \ref{fig:SourceSetup}, a fusion spliced, polarization maintaining pump signal combiner, the Yb PM 30/400 fiber under test, a cladding light stripper, and a fusion spliced angled end cap. PSC: pump signal combiner; PM: polarization maintaining; CLS: cladding light stripper; AR: anti-reflection.}
\label{fig:PowerAmp}
\end{figure*}

The power amplifier was constructed from a fusion spliced, 6+1:1 PM pump signal combiner (PSC, Lightera PN 7000698B), the Yb 30/400 fiber under test, a cladding light stripper (CLS) integrated into the gain fiber, and a fusion spliced anti-reflection (AR) coated angled end cap as show in Fig. \ref{fig:PowerAmp}. The PM 20/400 output fiber from the PSC was spliced directly to the PM Yb 30/400 without a mode-field adapter. The gain fiber was potted in the groove of a spiral plate with a pitch of 960\,$\upmu$m. In the following experiments, the inner diameter of the spiral was approximately 18 cm, and the outer diameter was around 18.5~cm. The etched region of the gain fiber that formed the CLS was around 6~cm in length, starting around 4~cm past the stripped coating edge of the fiber and ending around 3~cm before the endcap. The amplifier was pumped by up to 1600~W of power from 976~nm, wavelength locked diodes lasers, each emitting  \textasciitilde 65~W (nLIGHT e06). Four banks of diode lasers were utilized. In each bank, the diode lasers were coupled together using a pump combiner (Lightera CoolMode 7:1 pump combiner, part number 7000626B) into 230\,$\upmu$m core fiber. The outputs of the pump combiners were fusion spliced to the matching fiber pump input on the pump-signal combiner.

\begin{figure*}
\centerline{\includegraphics[width=6.5in]{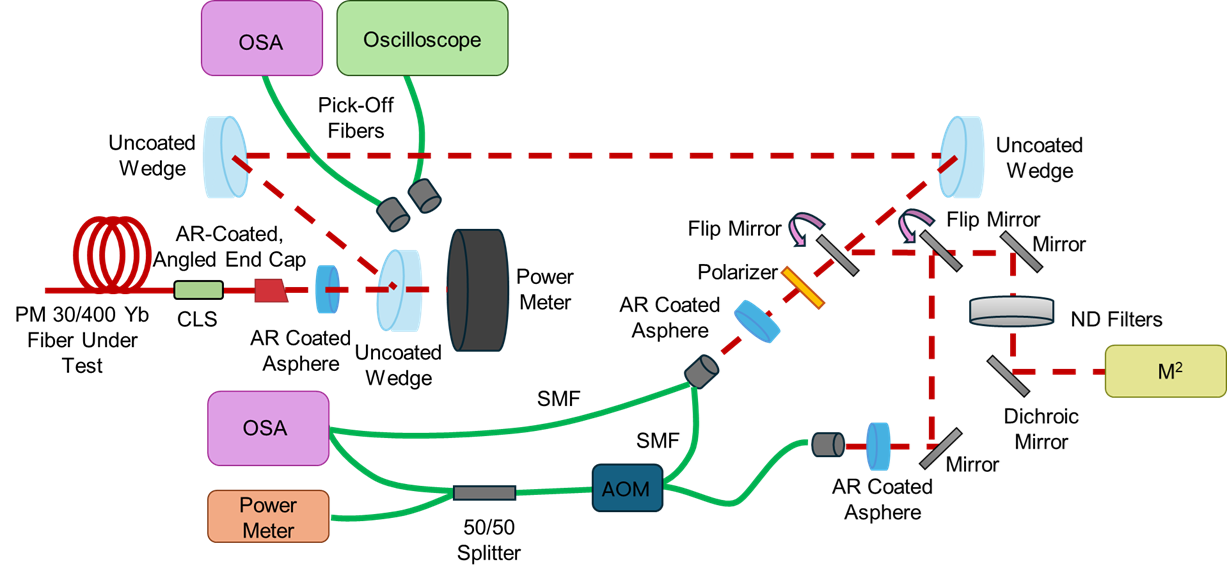}}
\caption{Depiction of diagnostics used to characterize the power amplifier performance as described in the main text. CLS: cladding light stripper; AR: anti-reflection; OSA: optical spectrum analyzer; SMF: single mode fiber; ND: neutral density; AOM: acousto-optic modulator.}
\label{fig:Diagnostics}
\end{figure*}

\begin{figure*}
\centerline{\includegraphics[width=5.5in]{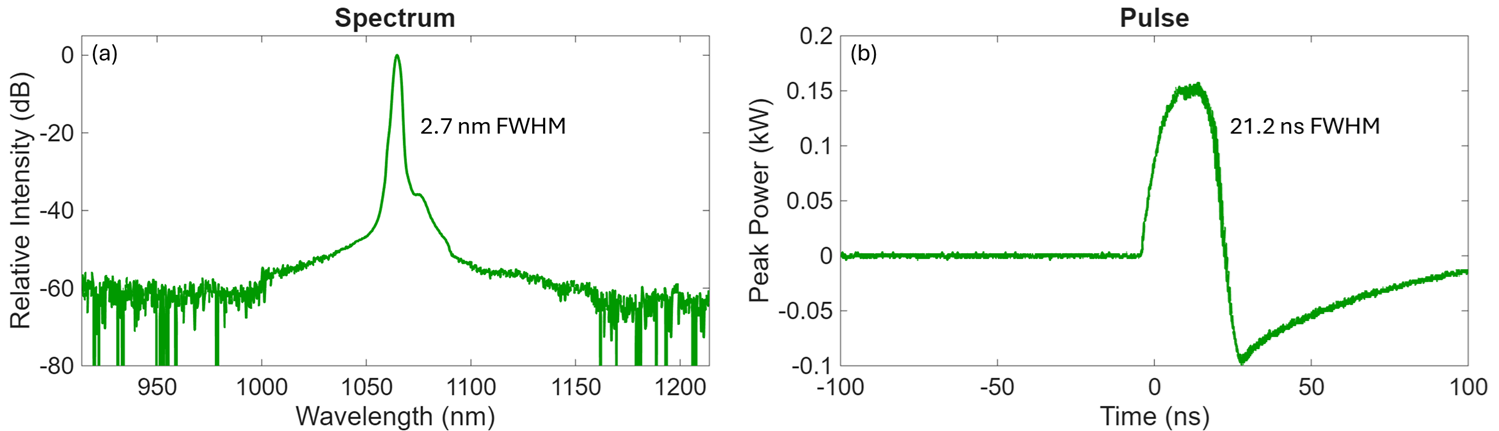}}
\caption{Seed pulse operating at 1~MHz and 3~W of average power (pulse energy of 3~\,$\upmu$J) showing the (a) spectrum and (b) pulse taken before the final polarizer prior to the power amplifier.}
\label{fig:SeedPulse}
\end{figure*}

The setup used to characterize the power amplifier is shown in Fig. \ref{fig:Diagnostics}. The output from the amplifier was collimated using an AR coated aspheric lens. The first of three uncoated wedges that were used to reduce the power prior to the majority of the diagnostics was placed between the collimation lens and the power meter used to measure the output power. Scatter from the power meter head is coupled to fibers and detectors and was used to measure the output spectrum, the pulse, and the pulse train. Light reflected from the third uncoated wedge could be sent to three different diagnostic paths through the use of two flip mirrors. The first path sent the light through a motorized polarizer, an AR coated aspheric lens, and back into single mode fiber (SMF) to measure the PER or the in-pulse energy of the polarized light through an acousto-optic modulator (AOM) measurement. The second beam path also coupled light into SMF via an AR coated aspheric lens to monitor the in-pulse energy of the light without the polarizer. Finally, a third beam path routed light to an M$^2$ measurement device (ThorLabs M2MS with BC207VIS), with a series of neutral density filters in the beam path that reduced the intensity to the appropiate level for the camera used to measure the M$^2$.

\section{Results}

\subsection{1~mJ Pulses at 1~kW}

\begin{figure*}
\centerline{\includegraphics[width=5.5in]{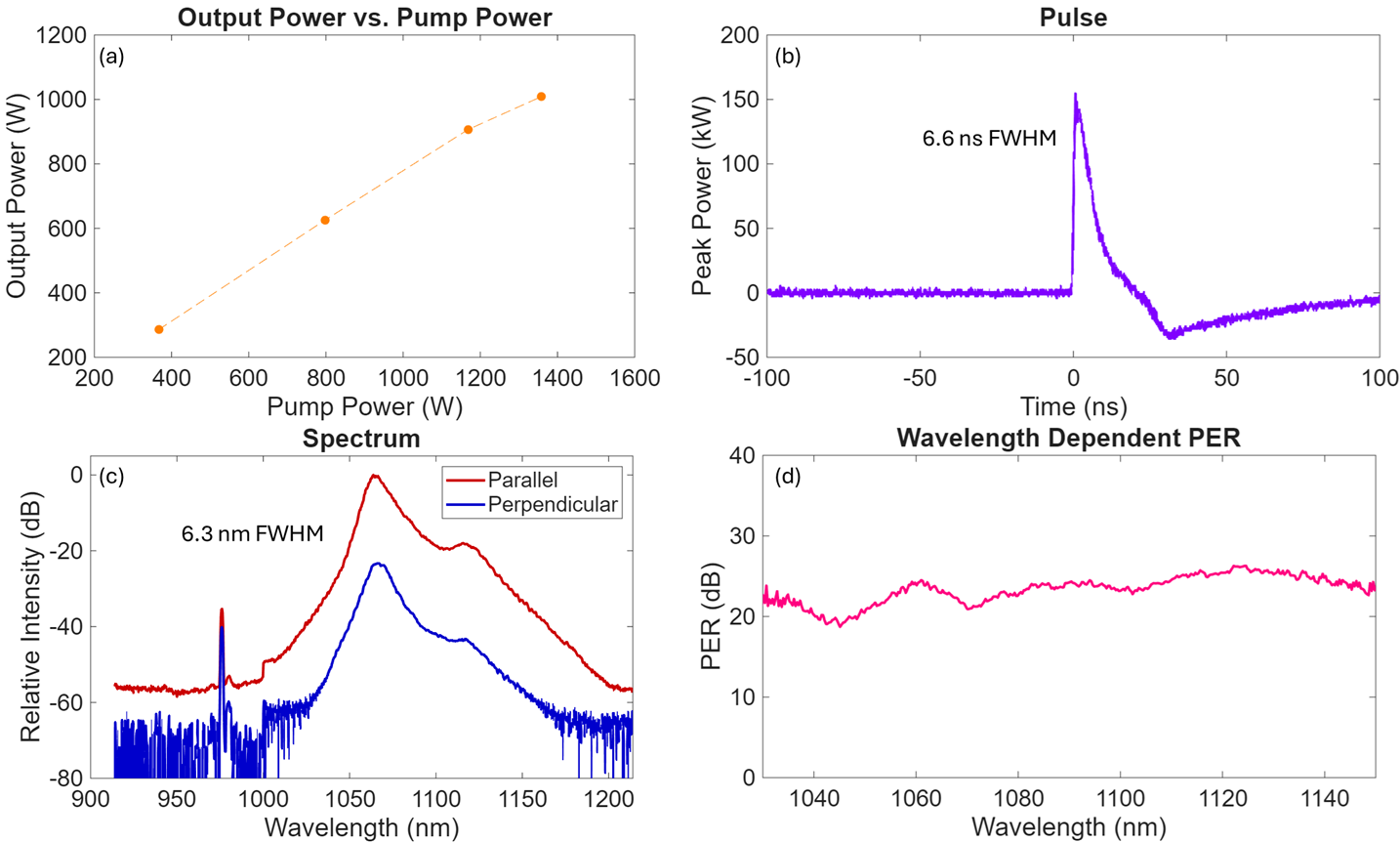}}
\caption{Power amplifier performance when operated at 1~MHz repetition rate with output pulse energy of 1~mJ and average power of 1~kW. (a) Output power vs. pump power. (b) Output pulse. (c) Output spectra taken after a polarizer parallel and perpendicular to the seed laser polarization axis. The Raman light at 1100~nm is 20~dB below the main peak. (d) Wavelength dependent polarization extinction ratio calculated by subtracting the spectra in (c).}
\label{fig:1mJ1kW}
\end{figure*}

The output spectrum and pulse from the seed system when operated at 3~W average power and 1~MHz repetition rate are shown in Fig. \ref{fig:SeedPulse}. The results from the power amplifier seeded with this pulse and operated at 1~kW of average power and 1~mJ of pulse energy are shown in Fig. \ref{fig:1mJ1kW}. The optical-to-optical efficiency, shown in Fig. \ref{fig:1mJ1kW}(a), was 74\% at the highest output power due to the fact that the fourth bank of pump diodes was not fully wavelength locked at that power level. At the peak performance of the wavelength-locked prior pump banks, the efficiency of the amplifier was 77.6\%. The seed pulse was 21.2~ns in duration, and the gain-shortened output pulse of 6.6~ns is shown in Fig. \ref{fig:1mJ1kW}(b). The shape of these pulses is typical of high gain amplifiers in this regime as the leading edge of the pulse depletes the gain medium and sees higher gain than the trailing edge \cite{Khitrov2008,Lin2017,Ran2015}. The amplified pulses had a peak power of 150~kW, and the 3~dB bandwidth of the spectrum was 6.3~nm. 

\begin{figure*}
\centerline{\includegraphics[width=5.5in]{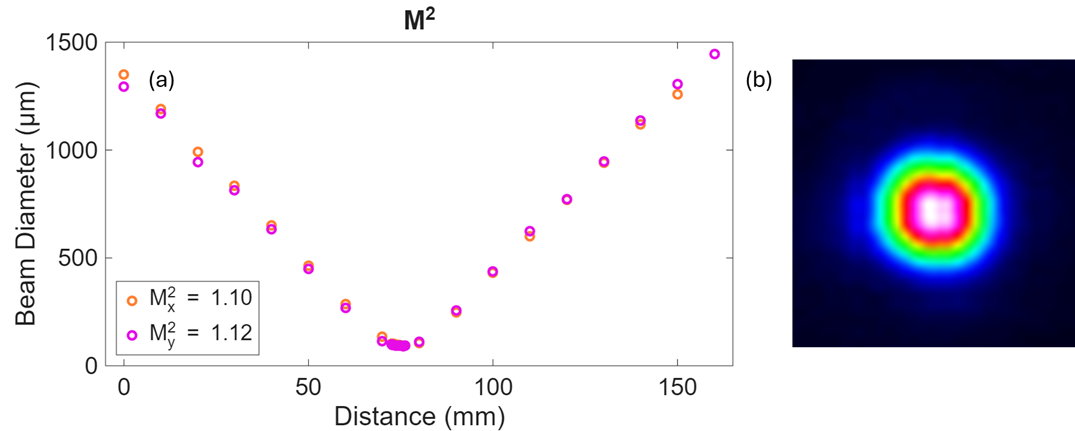}}
\caption{(a) Results of M$^{2}$ measurement and (b) the corresponding 2D beam profile at the focus of the M$^{2}$ measurement for the 1~mJ pulses.}
\label{fig:1mJ1kWM2}
\end{figure*}

The PER of the output pulses was measured by placing a polarizer after a series of reflections from uncoated wedges and recording the on-axis spectrum with the polarizer aligned parallel and perpendicular to the seed laser polarization axis. The wedges used to reduce the output power were operated at a low angle of incidence less than 15 degrees to minimize polarization dependent reflections. The resulting spectra are shown in Fig. \ref{fig:1mJ1kW}(c), and the subtraction of them is shown in Fig. \ref{fig:1mJ1kW}(d) with an average PER of 23.3~dB over this wavelength range. After a dichroic mirror to remove the residual pump light, the spatial beam quality was assessed using an M$^{2}$ measurement as shown in Fig. \ref{fig:1mJ1kWM2}, yielding 1.10 in the x-axis and 1.12 in the y-axis. 

\begin{figure*}
\centerline{\includegraphics[width=5.5in]{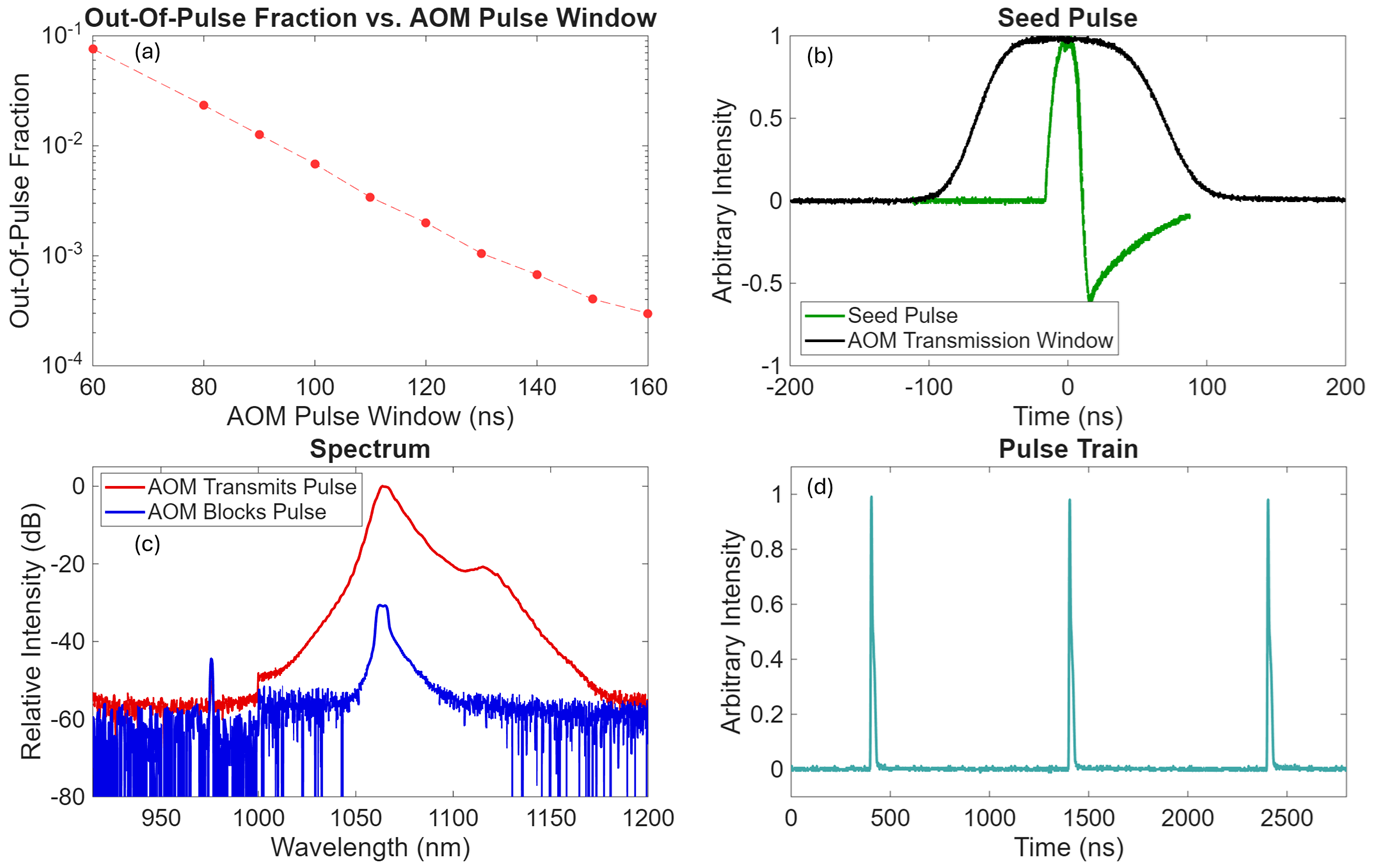}}
\caption{(a) Fraction of out-of-pulse energy measured with the 3~W seed pulse as a function of acousto-optic modulator (AOM) transmission window width. (b) Seed pulse in comparison to the AOM transmission window. (c) AOM measurement of in-pulse (red) versus out-of-pulse (blue) power of the 1~mJ pulses. (d) DC-coupled pulse train.}
\label{fig:1mJ1kWQuality}
\end{figure*}

Recent theoretical work has investigated the maximum energy extraction from LMA fiber amplifiers \cite{Bingham2025}. In order to ensure the measured power is representative of the coherent pulsed power and not ASE or CW power at the signal wavelength, we investigated the in-pulse power through the use of an AOM (NEOS 23080-1-1.06-LTD) as a time gate \cite{Headley2005} and through the use of a DC coupled photodiode. The measured suppression of the AOM was over 35~dB. The transmission window of the AOM was determined by monitoring the out-of-pulse energy fraction as a function of the AOM transmission window width as shown in Fig. \ref{fig:1mJ1kWQuality}(a). The transmission window of the AOM was determined by monitoring the out-of-pulse energy fraction of the seed pulse as a function of the AOM transmission
window width as shown in Fig. \ref{fig:1mJ1kWQuality}(a). The measurement was performed with the 150~ns window, at which point the out-of-pulse fraction of the seed pulse had plateaued at 0.04\% with respect to window width. The selected AOM window with respect to the seed pulse is shown in Fig. \ref{fig:1mJ1kWQuality}(b) and indicates an appropriate window duration given the 60~ns rise and fall time of the AOM relative to the seed pulse's full width at half maximum of 21.2~ns. The same AOM window length was used at all levels of amplification, so the window length was selected with respect to the seed pulse rather than the 1~mJ amplified pulse, as the amplified pulse duration depends on the amount of gain. The recorded in-pulse (red) and out-of-pulse (blue) spectra of the polarized pulse are shown in Fig. \ref{fig:1mJ1kWQuality}(c), and the corresponding power meter measurement showed that 99.94\% of the power resides in-pulse. This measurement repeated on the output pulse prior to the aligned polarizer yields a nearly identical result. The DC-coupled pulse train (recorded with a ThorLabs PDA10CF detector) is shown in Fig. \ref{fig:1mJ1kWQuality}(d) and shows no time-dependent background between pulses, indicating no ASE buildup or amplification of in-band signal wavelengths between the pulses, which would be evident as a increase in power prior to the arrival of the subsequent pulse.

\subsection{1.5~mJ Pulses at 750~W}

\begin{figure*}
\centerline{\includegraphics[width=5.5in]{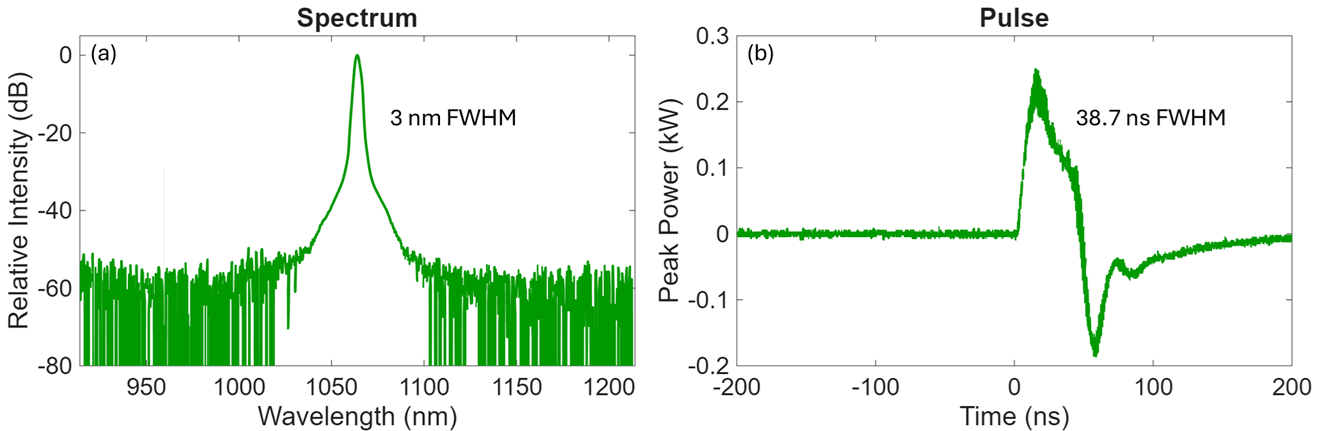}}
\caption{Seed pulse operating at 500~kHz and 3~W of average power (pulse energy of 6~\,$\upmu$J) showing the (a) spectrum and (b) pulse taken before the final polarizer prior to the power amplifier.}
\label{fig:1pt5mJSeed}
\end{figure*}

\begin{figure*}
\centerline{\includegraphics[width=5.5in]{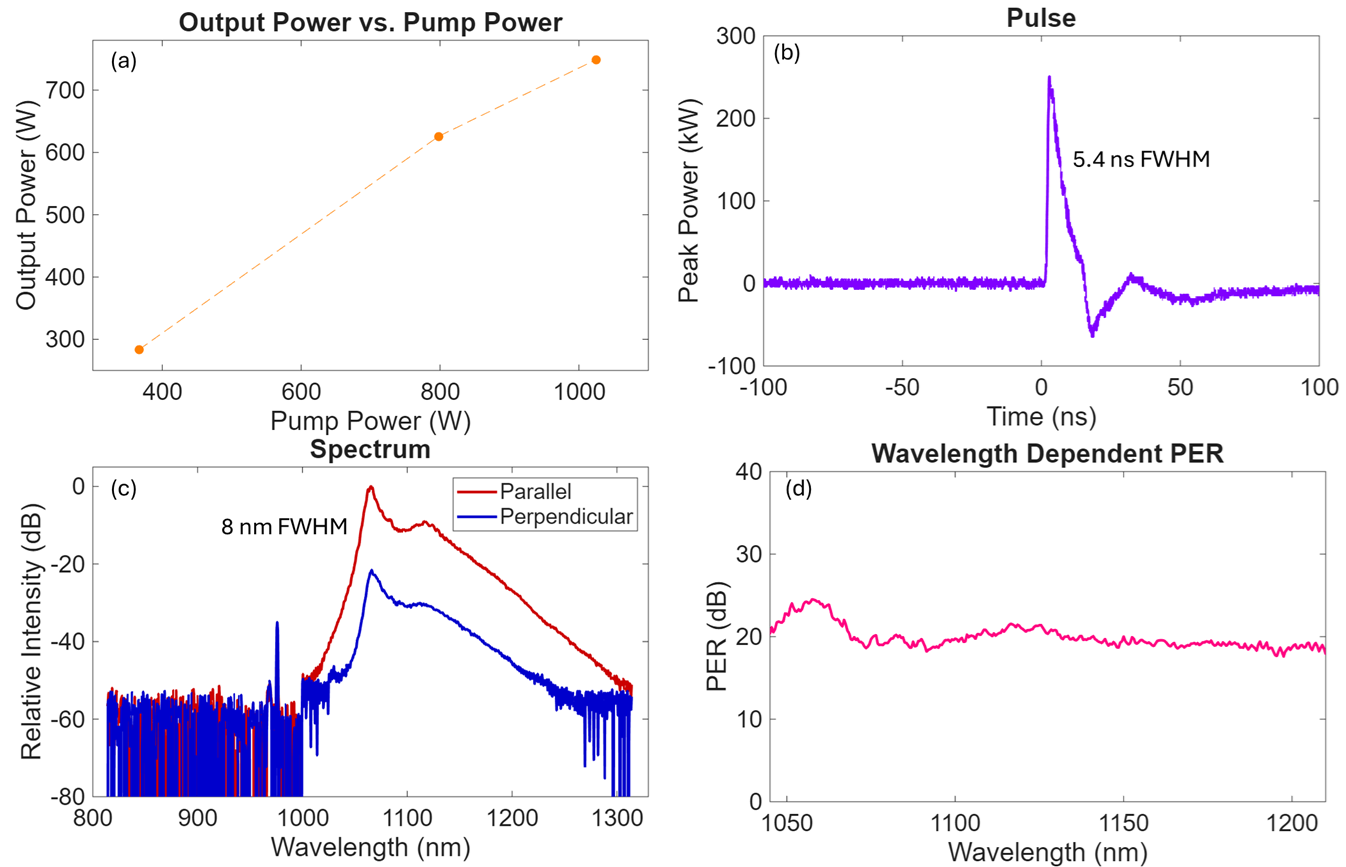}}
\caption{Power amplifier performance when operated at 500~kHz repetition rate with output pulse energy of 1.5~mJ and average power of 750~W. (a) Output power vs. pump power. (b) Output pulse. (c) Output spectra taken after a polarizer aligned parellel and perpendicular to the seed laser polarization axis. The Raman peak at 1100~nm is 10~dB below the main peak. (d) Wavelength dependent polarization extinction ratio calculated by subtracting the spectra in (c).}
\label{fig:1pt5mJ}
\end{figure*}

\begin{figure*}
\centerline{\includegraphics[width=5.5in]{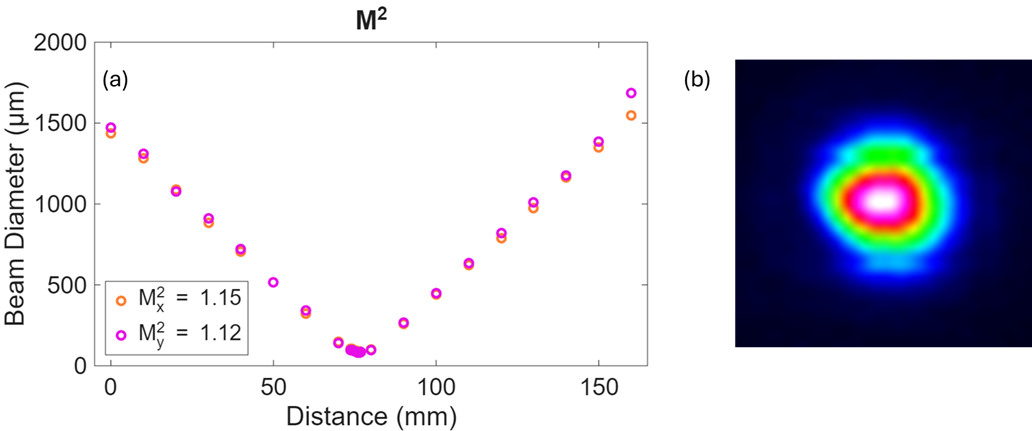}}
\caption{(a) Results of M$^{2}$ measurement and (b) the corresponding 2D beam profile at the focus of the M$^{2}$ measurement for the 1.5~mJ pulses.}
\label{fig:1pt5mJM2}
\end{figure*}

To further increase the pulse energy, the seed pulse length was increased to 38.7~ns, as shown in Fig. \ref{fig:1pt5mJSeed}(b), to reduce nonlinear effects, primarily SRS and SPM, by lowering the peak power for a given pulse energy. In addition, the repetition rate was lowered to 500~kHz to enable higher pulse energy while staying well below the TMI limit.  With these changes, the output pulse energy of the amplifier was able to reach 1.5 mJ. The performance of the amplifier with these modifications is shown in Fig. \ref{fig:1pt5mJ}. The optical-to-optical efficiency was 78.4\% at the peak efficiency of the second pump diode bank and 73\% at the maximum pulse energy. The 38.7~ns seed pulse shortened to 5.4~ns in the gain fiber and yielded a peak power of 250~kW. The output spectrum showed a higher Raman peak due to the higher pulse energy, and the average PER over the wavelength range shown in Fig. \ref{fig:1pt5mJ}(d) was 20~dB. The M$^{2}$ measurement indicated nearly diffraction limited beam quality with a result of 1.15 in the x-direction and 1.12 in the y-direction, as shown in Fig. \ref{fig:1pt5mJM2}.

\begin{figure*}
\centerline{\includegraphics[width=5.5in]{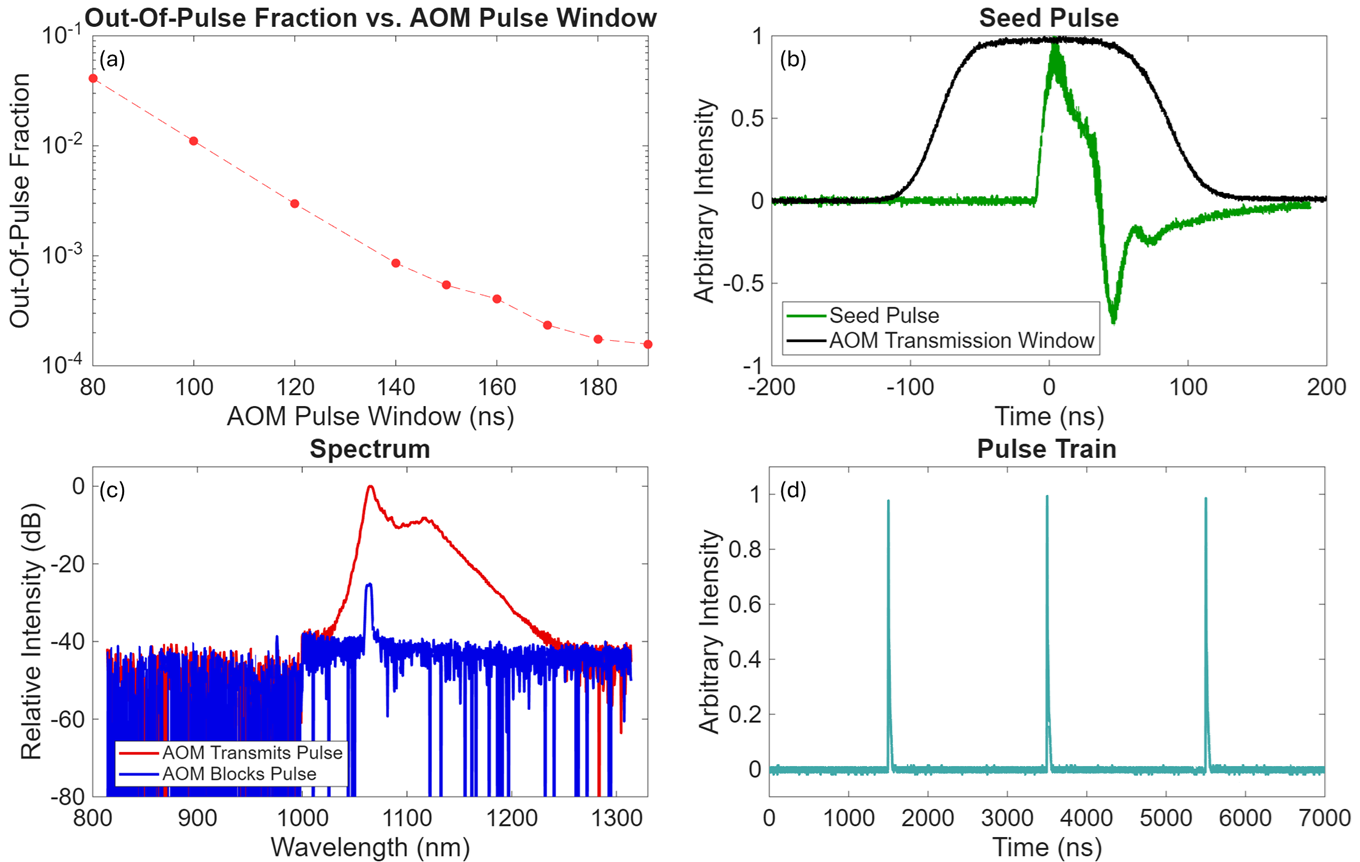}}
\caption{(a) Fraction of out-of-pulse energy measured with the 3~W seed pulse as a function of acousto-optic modulator (AOM) transmission window width. (b) Seed pulse in comparison to the AOM transmission window. (c) AOM measurement of in-pulse (red) versus out-of-pulse (blue) power of the 1.5~mJ pulses. (d) DC-coupled pulse train.}
\label{fig:1pt5mJQuality}
\end{figure*}

\begin{figure}
\centerline{\includegraphics[width=3.5in]{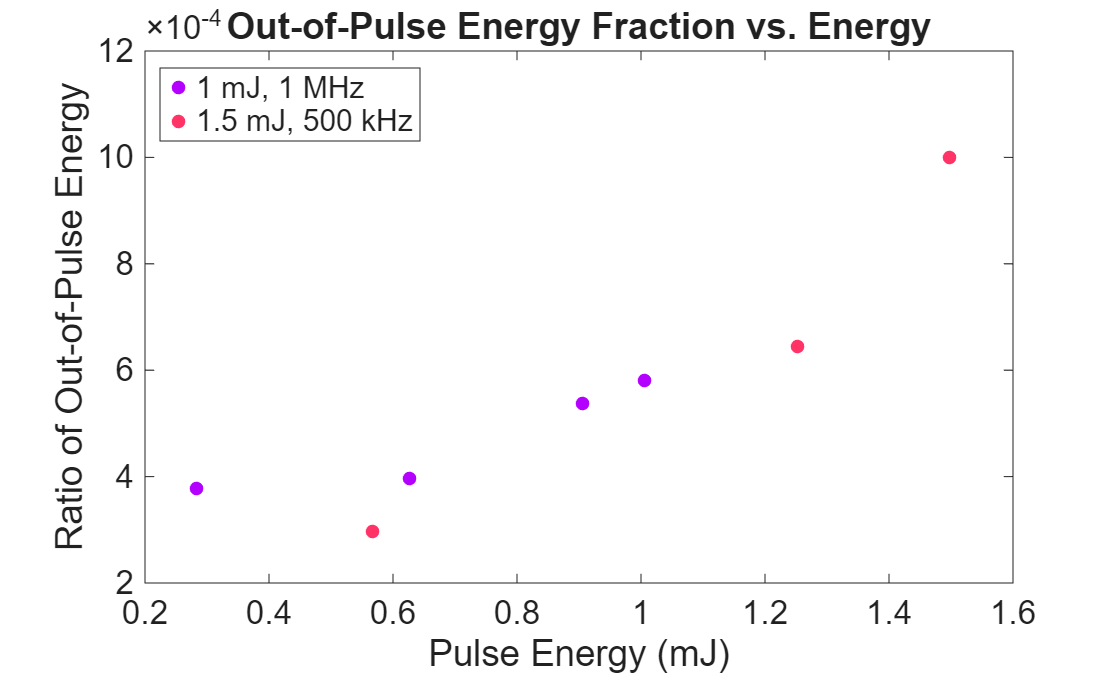}}
\caption{Comparison of out-of-pulse energy fraction measured via the acousto-optic modulator versus amplifier output energy.}
\label{fig:InPulseVsEnergy}
\end{figure}

The in-pulse energy measurement was verified as before, with a DC-coupled pulse train measurement and an AOM measurement comparing the in-pulse to out-of-pulse power, taken after the polarizer, and shown in Fig. \ref{fig:1pt5mJQuality}. For this amplifier configuration, the AOM transmission window was expanded to 180~ns to compensate for doubling the seed pulse duration and was determined as before, with a measured 0.02\% of the seed power out-of-pulse. The pulse train remained flat between successive pulses, and the power meter measurement that corresponded to the AOM measurement indicated 99.9\% of in-pulse power. However, Fig. \ref{fig:InPulseVsEnergy} indicates that the fraction of out-of pulse energy is increasing at a faster rate at the higher pulse energies, although even at the highest energy level, 99.9\% of the energy remains in the pulses. Thus, it will be important to continue monitoring this metric as pulse energies from these fibers are increased, with the associated increase in ASE at higher levels of gain.

\section{Discussion and Conclusion}

The 1~mJ and 1.5~mJ pulses achieved with the 26~\,$\upmu$m core PM fiber show that PM fibers can achieve similar performance to their non-PM counterparts, and that the enabling high HOM loss fiber designs are compatible with panda PM fibers. The high HOM loss, LMA fibers allow for pulsed amplification in a new regime of high average power and pulse energy, which can enable new applications such as long range coherent LIDAR with an all-fiber, environmentally stable source. The characterization presented in this work shows that at high repetition rates, power between pulses remains extremely low. Thus, further energy scaling in this fiber may be possible through additional shaping of the seed pulse. 

\bibliographystyle{elsarticle-num} 
\bibliography{PMPUlsedOFT}

\end{document}